\documentclass[preprint,12pt]{elsarticle}

\usepackage{amssymb}

\usepackage{ulem}

\newcommand{\dfrac}[2]{\displaystyle\frac{#1}{#2}}

\usepackage[usenames]{color}

\journal{Physics Letters A}

\makeatletter
\def\ps@pprintTitle{%
 \let\@oddhead\@empty
 \let\@evenhead\@empty
 \def\@oddfoot{}%
 \let\@evenfoot\@oddfoot}
\makeatother

\begin{document}

\begin{frontmatter}



\title{Circular symmetry 
in the Hitchin system}
\author{Masaru Kamata
}
\ead{kamata@kisarazu.ac.jp
}
\address{Natural Science Education, Kisarazu National College of Technology\\
2-11-1 Kiyomidai-Higashi, Kisarazu, Chiba 292-0041, Japan}

\begin{abstract}
To study circularly symmetric field configurations in the $SU(2)$ Hitchin system 
an $SO(2)$ symmetry, $[J_3, \phi]=0$ and $[J_3, A_{\pm}]=\pm A_{\pm}$, is imposed 
on the Higgs scalar $\phi$ and the gauge fields $A_{\pm}$ of the system, respectively, 
where $J_3$ is a sum of the third components of the orbital angular momenta 
and the generators of the $SU(2)$. 
The circular symmetry and the equation $\bar{D}\phi=0$ 
yield constant, generally nonzero, vacuum expectation values for 
${\rm Tr}(\phi^{2})$. The equation $4F_{z\bar{z}}=[\phi,\,\phi^{*}]$ 
yields a system of differential equations which govern the circularly symmetric field 
configurations and an exact solution to these equations 
in a pure gauge form with nontrivial Higgs scalar is obtained.
\end{abstract}

\end{frontmatter}


\section{Introduction}
The BPST instantons \cite{BPST} in the Euclidean four-dimensional space $\mathbb{R}^4$ and the BPS 
monopoles \cite{BogoPS} in the three-dimensional space $\mathbb{R}^3$ are well described and constructed 
through the ADHM \cite{AHDM} and Nahm \cite{Nahm80, Nahm82} constructions, respectively. 
However, for the Hitchin system \cite{Hit86, Lohe} which is defined on the two-dimensional 
space $\mathbb{R}^2$, physically interesting solutions have not been known so far. 
Although the system has a noble property, {\it i.e.}, the codimension of the space $\mathbb{R}^2$ 
in $\mathbb{R}^4$ is also two and the system is at a fixed point of 
the reciprocity \cite{CorriGodd, Fields Medal Symposium},
the analysis on the system is not yet sufficient both from the mathematical and physical points of view.

It is well known \cite{COFN} for the spherically symmetric 'tHooft-Polyakov magnetic 
monopole \cite{tH, Polyakov} that the Higgs scalar $\phi$ and the gauge potentials $A_{j}$ 
are subject to the conditions 
\begin{equation}
[J_{j}, \phi]=0, \:\: [J_{j}, A_{k}]=i\varepsilon_{jkl}A_{l}, \label{tHP SO(3)sym} 
\end{equation}
where $j,\:k$ and $l$ run from 1 to 3 and $J_{j}=L_{j}+T_{j}$ are the sum of the orbital angular 
momenta $L_j$ and the generators $T_j$ of $SU(2)$ which is locally isomorphic to $SO(3)$. 
The Higgs scalar $\phi$ and the gauge potentials $A_{j}$ transform as a scalar and a vector 
under the subgroup $SO(3)_{\bf L+T}$ of $SO(3)_{\bf L}\otimes SO(3)_{\bf T}$, respectively. 
The aim of this letter is to study circularly symmetric 
field configurations in the $SU(2)$ Hitchin system and to examine their properties. 
For this end we will impose an $SO(2)_{J_3}$ symmetry 
\begin{equation}
[J_3, \phi]=0, \:\: [J_3, A_{\pm}]=\pm A_{\pm}, \label{[J phi]=0, [J A]=A} 
\end{equation}
on Hitchin's scalar $\phi$ and the gauge potentials $A_{\pm}=A_1\pm iA_2$, respectively. 
The conditions (\ref{[J phi]=0, [J A]=A}) are obtained from (\ref{tHP SO(3)sym}) 
by restricting the range of indices to $j=3$ and $k, \:l=1, \:2$ and the $\pm$ signs represent 
the helicity of $A_{\pm}$. Here we have defined 
$J_3:=L_3+iT_3$, where the imaginary unit $i$ is due to our convention used in this letter. 
We will show that the circular symmetry (\ref{[J phi]=0, [J A]=A}) and the equation 
$\bar{D}\phi=0$ of the Hitchin system yield constant ${\rm Tr}(\phi^2)$, whose value is 
determined by the asymptotic values of the Higgs scalar at the infinity and generally nonzero. 
This circumvents Derrick's theorem \cite{MantonSut} or 
a dimensional argument \cite{CherkisKap} about the nonexistence 
of finite action configurations in the system. 
The interpretation of these circularly symmetric configurations 
as topologically stable Abrikosov-Nielsen-Olesen vortices \cite{NO} fails 
because the surface term in the action integral 
vanishes for these configurations.  
Finally an exact solution in a pure gauge form with nontrivial Higgs scalar will be presented. 
Although the field strength has a pure gauge form $F_{z\bar{z}}=0$, the solution is not 
a genuine pure gauge one, 
because the relevant gauge transformation is singular. 

The letter is organized as follows: 
In Section 2 we will briefly summarize the Hitchin system. 
In Section 3 we will impose the circular symmetry on the Hitchin system and see 
that ${\rm Tr}(\phi^{2})$ is a constant. 
In Section 4 we will derive differencial equations for circularly symmetric configurations. 
In Section 5 an exact solution in a pure gauge form with nontrivial Higgs scalar will 
be presented. 
Section 6 will be devoted to conclusions and outlook.

\section{The Hitchin system}

We here briefly summarize the Hitchin system \cite{Hit86, Lohe}.  
This system is derived by imposing a two-dimensional translational invariance 
on the four-dimensional self-dual (SD) Yang-Mills equations $F=\!\,^{*}F$.  
Assuming the system to be invariant under two translations $\partial_3=\partial_4=0$, 
we have a complementary two-dimensional $x_1x_2$ space 
and obtain two Higgs scalars $\phi_1=A_3$ and $\phi_2=A_4$ defined on this plane. 
In our convention forms are anti-hermetian; a 1-form $A$, for example, is defined by 
$A=\sum^3_{a=1}\sum^4_{\mu=1}T_aA^{a}_{\mu}dx^{\mu}$, 
where $T_a=\frac{\sigma_a}{2i}$ are anti-hermetian $SU(2)$ generators and $\sigma_a$ 
are the Pauli matrices. The SD Yang-Mills equations $F=\!\,^{*}F$ reduce to 
  \begin{eqnarray}
  &4F_{z\bar{z}}=[\phi,\,\phi^{*}], \label{eq:1st Hitchin} \\ 
  &\bar{D}\phi=[\nabla_1+i\nabla_2,\phi]=0,\,\,\,D\phi^*=[\nabla_1-i\nabla_2,\phi^*]=0, \label{eq:2nd Hitchin}
  \end{eqnarray}
where $z=x_1+ix_2$, and $\nabla_j=\partial_j+A_j$ are covariant derivatives
with respect to the two-dimensional gauge potentials. 
The complex Higgs scalar $\phi$ and its conjugate $\phi^*$ are defined by
$\phi=\phi_1-i\phi_2$ and $\phi^*=\phi_1+i\phi_2$, respectively. 
Note that $\phi^{\dagger}=-\phi^*$ and $(\bar{D}\phi)^{\dagger}=-D\phi^*$, 
where the symbol $\dagger$ represents the hermitian conjugation and the minus signs 
in the right hand sides are due to anti-hermiticity of our $T_j$. 

\section{Circularly symmetric field configurations}

Let us consider hereafter the case of plus helicity in the condition (\ref{[J phi]=0, [J A]=A}). 
Using the polar coordinates $(\rho, \theta)$ on $\mathbb{R}^2$ with $\rho=\sqrt{x^2_1+x^2_2}\;$ 
and $\theta$ the polar angle, we have $L_3=-i\partial_{\theta}$ and the $\theta$ dependence of 
$\phi$ and $A_{+}$ can be determined from (\ref{[J phi]=0, [J A]=A}) as follows
\begin{equation}
\phi=
   -i\left(
   \begin{array}{ccccccc}
   h              &f_2e^{-i\theta} \\
   f_1e^{i\theta}&-h  
   \end{array}\right), 
\,\,\, A_{+}=
   -i\left(
   \begin{array}{ccccccc}
   a &be^{-i\theta} \\
   ce^{i\theta} &-a  
   \end{array}\right)e^{i\theta}, \label{phi,A_{+}}   
\end{equation}
where $f_j \:\:(j=1, \:2)$, $h$, $a$, $b$ and $c$ are functions of $\rho$ only. 
Note that $A_{+}$ has the same $\theta$ dependence as $\phi$ except for the 
overall extra factor $e^{i\theta}$; 
this is because $\phi$ and $A_{+}$ are transformed as scalar and vector, respectively, 
under the $SO(2)_{J_3}$ rotation.  
The Higgs scalar $\phi=\phi(\rho,\theta)$ of (\ref{phi,A_{+}}) can be expressed as
\begin{equation}
\phi=g(\theta)\phi_0g(\theta)^{-1} \label{gphi_0g^{-1}}, 
\end{equation}
where $\phi_{0}=\phi|_{\theta=0}$ and $g(\theta)={\rm exp}(T_3\theta) \in SU(2)$. 
We can remove the factor $g(\theta)$ of (\ref{gphi_0g^{-1}}) 
by singular gauge transformation $g(\theta)^{-1}$, 
but this gives rise to a two-dimensional delta function singularity $2\pi\delta^2(x)T_3$ 
\cite{Kamata1978}, a Dirac string term \cite{Dirac monopole}, to $F_{z\bar{z}}$. 
We will discuss this point again later. 

Since $\phi$ is traceless, we have from (\ref{gphi_0g^{-1}}) 
\begin{equation}
\phi^{2}=-{\rm det}\phi\cdot 1_2=-{\rm det}\phi_0\cdot 1_2 \label{phi^2=det}.
\end{equation}
The equation $\bar{D}\phi=0$ of (\ref{eq:2nd Hitchin}) implies $\bar{D}(\phi^2)=0$, 
which yields $\partial_{\bar{z}}{\rm det}\phi_0=0$ because of (\ref{phi^2=det}). 
In the polar coordinates, this equation is expressed as 
$e^{i\theta}(\partial_{\rho}+\frac{i}{\rho}\partial_{\theta}){\rm det}\phi_0=0$ 
and we have $\partial_{\rho}{\rm det}\phi_0=0$. 
Therefore we see that $\det{\phi}=f_1f_2+h^2$ is a constant, whose value 
is determined by the asymptotic values of $f_1,\:f_2$ and $h$ at the infinity and is 
generally nonzero. 
The significant point here is that the constant valuedness of 
${\rm Tr}(\phi^{2})=-2(f_1f_2+h^2)$ is a consequence of the dynamics and symmetry of 
the system. This situation is quite different from other typical models in mathematical 
physics, for example, in non-linear sigma models, the constraint on the scalar 
fields $|\vec{\phi}|=1$ is imposed by hand, or implemented through the Fayet-Iliopoulos 
term, which will introduce additional parameters into the models. 

\section{Differencial equations for circularly symmetric configurations} 

Substituting (\ref{phi,A_{+}}) for $\phi$ and $A_{+}$ of 
$\bar{D}\phi=0$ of (\ref{eq:2nd Hitchin}) we have 
\begin{eqnarray}
i\Bigl(\dfrac{df_{1}}{d\rho}-\dfrac{f_{1}}{\rho}\Bigr)+2(ch-af_{1})&=&0, \label{f'_{1}} \\
i\Bigl(\dfrac{df_{2}}{d\rho}+\dfrac{f_{2}}{\rho}\Bigr)+2(af_{2}-bh)&=&0, \label{f'_{2}} \\
i\dfrac{dh}{d\rho}+bf_{1}-cf_{2}&=&0, \label{h'}
\end{eqnarray}
from which we again see that ${\rm det}\phi=f_1f_2+h^2$ is a constant.  
Substituting (\ref{phi,A_{+}}) for (\ref{eq:1st Hitchin}) we have the following system of 
differential equations 
\begin{eqnarray}
-2\Bigl(\dfrac{d}{d\rho}+\dfrac{1}{\rho}\Bigr){\rm Im}(a)+|c|^2-|b|^2&=&|f_{2}|^2-|f_{1}|^2, \label{Diag} \\
i\Bigl(\dfrac{d}{d\rho}+\dfrac{2}{\rho}\Bigr)c-i\dfrac{d\bar{b}}{d\rho}+2(\bar{b}a-\bar{a}c)&=&2(\bar{h}f_{1}-\bar{f_{2}}h), \label{Off-diag} 
\end{eqnarray}
which stem from the diagonal and off-diagonal parts of (\ref{eq:1st Hitchin}), respectively. 

The linear equations (\ref{f'_{1}}), (\ref{f'_{2}}) and (\ref{h'}) for $a$, $b$ and $c$ 
are underdetermined system, because the gauge potential $A_{+}$ of the equation 
$\bar{D}\phi=2\partial_{\bar{z}}\phi+[A_{+},\phi]=0$ of (\ref{eq:2nd Hitchin})
is not unique. We observe that the general solution is given by 
\begin{equation}
A_{+}=\lambda \phi+\dfrac{1}{2h}
\left(
   \begin{array}{ccccccc}
   0              &\tilde{f}_2 \\
   -\tilde{f}_1e^{2i\theta} &0  
   \end{array}\right), \label{A_{+}}
\end{equation}
where $\lambda$ is an arbitrary complex scalar and $\tilde{f}_1$ and $\tilde{f}_2$ are 
defined by
\begin{equation}
\tilde{f}_1=f'_1-\dfrac{f_1}{\rho},\,\,\, \tilde{f}_2=f'_2+\dfrac{f_2}{\rho}, 
\end{equation}
respectively, in which the prime stands for the differentiation with respect to $\rho$. 
Similar circumstances occurred in a study of the $SU(3)$ magnetic monopoles by 
Corrigan et al.\,\cite{COFN}. They solved the equations $D_{\mu}\phi=0$ for the gauge 
potentials $A_{\mu}$ and obtained asymptotic field strengths $F_{\mu\nu}$ at infinities. 

A convenient choice of $\lambda$, or gauge, in (\ref{A_{+}}) is given by 
$\lambda=\frac{i\tilde{f_1}}{2f_1h}e^{i\theta}$, which corresponds to $c=0$. 
In this case, $A_{+}$ becomes an upper triangular matrix 
in the Atiyah and Ward form \cite{AtiyahWard} 
\begin{equation}
A_{+}=\dfrac{1}{2f_1}
\left(
   \begin{array}{ccccccc}
   \tilde{f_1}e^{i\theta}  & -2h' \\
   0                       & -\tilde{f_1}e^{i\theta}  
   \end{array}\right). \label{A_{bar{z}}}
\end{equation}
From this we see $a=\frac{i \tilde{f}_1}{2{f}_1}$ and $b=-\frac{ih'}{{f}_1}$ and 
the equations (\ref{f'_{1}}) - (\ref{Off-diag}) reduce to the equations 
\begin{eqnarray}
f_{1}f_{2}+h^2&=&C^2, \label{f_{1}f_{2}+h^2} \\
\Biggl(\dfrac{|f_{1}|'}{|f_{1}|}\Biggr)'+\dfrac{1}{\rho}\dfrac{|f_{1}|'}{|f_{1}|}+\Biggl|\dfrac{h'}{f_{1}}\Biggr|^2&=&|f_{1}|^2-|f_{2}|^2, \label{Diag2} \\
\Biggl(\dfrac{h'}{f_{1}}\Biggr)'+\dfrac{1}{\rho}\dfrac{h'}{f_{1}}-\dfrac{\bar{f'_{1}}h'}{|f_{1}|^2}&=&2(\bar{f_{1}}h-\bar{h}f_{2}), \label{Off-diag2}
\end{eqnarray}
where $C$ is a constant. Note that only the absolute value of $f_{1}$ appears 
in (\ref{Diag2}). If we expand $f_j\:(j=1, \:2)$ and $h$ in power series 
of $\rho$ around $\rho=0$ 
\begin{equation}
f_j=f_{j}^{(0)}+f_{j}^{(1)}\rho+f_{j}^{(2)}\rho^2+\cdots, \:\:\: h=h^{(0)}+h^{(1)}\rho+h^{(2)}\rho^2+\cdots, 
\quad \label{expansion rho=0}
\end{equation} 
we see $f_{j}^{(0)}=f_{j}(0)=0$ from the single-valuedness of $f_j$ at $\rho=0$. 
We also have $h^{(1)}=0$ from (\ref{f_{1}f_{2}+h^2}) when $C= \pm h^{(0)} \ne 0$. 
Using (\ref{expansion rho=0}) in (\ref{Diag2}) and (\ref{Off-diag2}) 
we will obtain a formal power series solution with coefficients fulfilling some 
relations between them. We have also asymptotic solution 
$f_1 \to f_{1}^{\infty}, \; f_2 \to f_{2}^{\infty}$ and $h \to h^{\infty}$
at the infinity $\rho \to \infty$ where $f_{1}^{\infty}, \: f_{2}^{\infty}$ and 
$h^{\infty}$ are constants satisfying 
the condition $f_{1}^{\infty}f_{2}^{\infty}+({h^{\infty}})^2=C^2$. 

\section{An exact solution in a pure gauge form
with nontrivial Higgs scalar}

We here consider a special case such that $f_{1},\;f_{2}$ and $h$ are all real and 
the relation $f_{1}=f_{2}\:(\equiv f)$ holds 
at some interval of $\rho$. In this case, since the right-hand sides of (\ref{Diag2}) 
and (\ref{Off-diag2}) vanish, we have $F_{z\bar{z}}=0$. 
Putting $t=\ln{\rho}$ and $\ln(\frac{dh}{dt}/{f})=-u$ in (\ref{Diag2}) and (\ref{Off-diag2})
we can derive a one-dimensional Liouville equation
\begin{equation}
\dfrac{d^2u}{dt^2}=e^{-2u} \label{eq:Liouville}
\end{equation}
with a minus sign in the exponent. A general solution to (\ref{eq:Liouville}) is given by 
$u=\ln\{{\nu}^{-1}\cosh\nu(t-t_0)\}$, 
where $\nu$ and $t_0$ are constants of integration and $f$ and $h$ are given by 
\begin{equation}
f=C\dfrac{2\nu\rho^{\nu}_0\rho^{\nu}}{\rho^{2\nu}+\rho^{2\nu}_0}, 
\:\: h=C\dfrac{\rho^{2\nu}-\rho^{2\nu}_0}{\rho^{2\nu}+\rho^{2\nu}_0}, \label{sol(f_1,h)}
\end{equation}
with $\rho_0=e^{t_0}$. Note that $f$ and $h$ of (\ref{sol(f_1,h)}) is a parameterization 
of an ellipse $f^2/{\nu}^2+h^2=C^2$. 
Without loss of generality we can restrict ourselves to the case $\nu >0$, 
because changing the sign $\nu \to -\nu$ in (\ref{sol(f_1,h)}) yields $f \to -f$ and $h \to -h$ 
which do not alter the equations (\ref{f_{1}f_{2}+h^2})-(\ref{Off-diag2}). 

If we apply the solution (\ref{sol(f_1,h)}) to the whole interval $0 \le \rho <\infty$ 
we have a solution in a pure gauge form with nontrivial Higgs scalar, 
in which $f$ tends to zero both as $\rho \to 0$ and as $\rho \to \infty$, and 
$h$ connects smoothly two boundary values $\pm C$ at the infinity $\rho \to \infty$ and 
at the origin $\rho=0$, respectively. 
Although the field strength in this case has a pure gauge form $F_{z\bar{z}}=0$, 
the solution (\ref{sol(f_1,h)}) is not a genuine pure gauge one, because the relevant 
gauge transformation is singular. In fact \cite{Kamata1978}, after performing the singular 
gauge transformation $g=g(\theta)^{-1}={\rm exp}(-T_3\theta)$, we have a nonzero field strength 
\begin{equation}
F^{g}_{12}=g(\theta)^{-1}F_{12}g(\theta)+g(\theta)^{-1}[\partial_{1}, \partial_{2}]g(\theta)=2\pi \delta^2(x)T_3, \quad \label{F^{g}}
\end{equation} 
which corresponds to a Dirac string term \cite{Dirac monopole}. 
We can show (\ref{F^{g}}) by explicit calculation. The transformed potentials are given by
\begin{eqnarray}
A_1^{g}&=&\dfrac{h'}{f_{1}}(T_1\sin{\theta}+T_2\cos{\theta})-\dfrac{\tilde{f}_{1}}{f_{1}}T_3\sin{\theta}+T_3\partial_1{\theta} , \label{A_1^{g}}\\
A_2^{g}&=&-\dfrac{h'}{f_{1}}(T_1\cos{\theta}-T_2\sin{\theta})+\dfrac{\tilde{f}_{1}}{f_{1}}T_3\cos{\theta}+T_3\partial_2{\theta}, \label{A_2^{g}}
\end{eqnarray}
and we have the field strength
\begin{eqnarray}
F_{12}^{g}&\!\!\!=\!\!\!&\partial_1A_2^{g}-\partial_2A_1^{g}+[A_1^{g}, A_2^{g}] \nonumber \\
          &\!\!\!=\!\!\!&-\Biggl\{\Biggl(\dfrac{h'}{f_{1}}\Biggr)'+\dfrac{1}{\rho}\dfrac{h'}{f_{1}}-\dfrac{{f'_{1}}h'}{f_{1}^2}\Biggr\}T_1 
          +\Biggl\{\Biggl(\dfrac{f_{1}'}{f_{1}}\Biggr)'+\dfrac{1}{\rho}\dfrac{f_{1}'}{f_{1}}+\Biggl(\dfrac{h'}{f_{1}}\Biggr)^2\Biggr\}T_3 +T_3[\partial_1, \partial_2]\theta \nonumber \\
          &\!\!\!=\!\!\!&T_3\nabla{\bf \cdot} \dfrac{\vec{\rho}}{\rho^2}=2\pi \delta^2(x)T_3,
\end{eqnarray}
where equations (\ref{Diag2}) and (\ref{Off-diag2}) have been used in the last line and then we have obtained (\ref{F^{g}}).
In this singular gauge 
the Higgs scalar has no $\theta$ dependence: 
$\phi'=g^{-1} \phi g=\phi_0$ where $\phi_0=\phi|_{\theta=0}$ of (\ref{gphi_0g^{-1}}).

\section{Conclusions and outlook}

We have found that the equation $\bar{D}\phi=0$ together with the circular 
symmetry (\ref{[J phi]=0, [J A]=A}) yields a constant, generally nonzero, 
vacuum expectation value for ${\rm Tr}(\phi^{2})$.
We have also seen that the equation $4F_{z\bar{z}}=[\phi,\,\phi^{*}]$ yields a system 
of differential equations and from which an exact solution in a pure gauge form with 
nontrivial Higgs scalar is obtained. Physical meaning of this solution is not clear and 
further study is needed, 
which will be discussed in a separate paper.

We finally discuss the surface terms in the action integrals which usually emerge 
when considering the SD, or the BPS, conditions. After a dimensional reduction from 
the four-dimensional Yang-Mills action, we consider the following action integral 
defined over a disk $D=\{(\rho, \theta)|\: 0 \le \rho \le R,\: 0 \le \theta \le 2\pi \}$ 
 \begin{eqnarray}
 S&=&\dfrac{1}{2}{\rm Tr}\int_D \{(F_{12})^2+\bar{D}\phi_1D\phi_1+\bar{D}\phi_2D\phi_2+[\phi_1,\phi_2]^2\}\,d^2x \nonumber \\
  &=&\dfrac{1}{16i}{\rm Tr}\int_D \{(4F_{z\bar{z}}-[\phi,\,\phi^{*}])^2+4(\bar{D}\phi)^{\dagger} \bar{D}\phi \bigr\}\,dzd\bar{z}+{\rm Tr}\int_D \partial_jJ_j\,d^2x, \qquad \label{S+sterm}
 \end{eqnarray}
where $J_j=-\varepsilon_{jk}{\rm Tr}\phi_1D_k\phi_2$ and 
$D_k\phi_2=\partial_k\phi_2+[A_k,\phi_2]$. 
The total divergence term in the last expression can be rewritten as a line integral 
along the boundary $\partial D=S^1_R$. For the circularly symmetric 
configurations (\ref{phi,A_{+}}), we have
\begin{eqnarray}
{\rm Tr}\int_D \partial_jJ_j\,d^2x
&=&{\rm Tr}\int_{S^1_R}(J_1\cos{\theta}+J_2\sin{\theta})\rho\,d\theta \nonumber \\
&=&-2\pi(|f_1|^2+|f_2|^2+2|h|^2)'\rho \Bigl|^R_0, \label{|^R_0}
\end{eqnarray}
which vanishes \cite{Lohe, MantonSut} as $R \to \infty$ due to boundary conditions 
discussed in Section 4. Since the boundary term (\ref{|^R_0}) vanishes the interpretation 
of these configurations as topologically stable Abrikosov-Nielsen-Olesen vortices \cite{NO} 
fails.  

Group theoretically, however, we expect the existence of topologically stable solitons. 
For the 'tHooft-Polyakov magnetic monopole, since the isotropy subgroup of the Higgs 
scalar $\phi$ is given by $\{{\rm exp}(\phi \alpha)| \alpha \in \mathbb{R} \} \cong SO(2)$, 
we have the second homotopy group 
$\pi_2(SO(3)/SO(2)) \cong \pi_2(S^2) \cong \mathbb{Z}$; 
this gives topologically stable magnetic monopoles. 
As for the Hitchin system, if $\phi_1$ and $\phi_2$ are not parallel nor anti parallel \cite{NO} 
then the isotropy subgroup of $\phi=\phi_1-i\phi_2$ is trivial and in this generic case we have 
the first homotopy group $\pi_1(SO(3)) \cong \pi_1(SU(2)/\mathbb{Z}_2) \cong \mathbb{Z}_2$.
Note that ${\rm exp}(\phi \alpha)$ does not belong to $SO(3)$ but to $SL(2, \mathbb{C})$. 
From this consideration, we expect the existence of topologically stable solitons in the 
circularly symmetric configurations in the Hitchin system. 
For example, we can extend the conditions (\ref{[J phi]=0, [J A]=A}) to more general ones
\begin{equation}
[J_3, \phi]=l\phi,\,\,\, [J_3, A_{\pm}]=\pm(l+1) A_{\pm}. \label{[J phi]=nphi, [J A]=(n+1)A} 
\end{equation}
Solving these we have $\phi=e^{il\theta}\phi^{(0)}$ and $A_{\pm}=e^{il\theta}A_{\pm}^{(0)}$, 
where $\phi^{(0)}$ and $A_{\pm}^{(0)}$ are the $\phi$ and the $A_{+}$ of (\ref{phi,A_{+}}), 
respectively. For $l=0$, the highest symmetry, we have seen that the surface term ($\ref{|^R_0}$) 
vanishes. For $l \ne 0$ we expect the existence of a field configuration 
with non-zero surface term, which will be examined elsewhere. \\

{\bf Acknowledgements}\\

The author would like to thank T. Hori, T. Koikawa, and S. Saito for helpful comments. 
He would like to thank especially A. Nakamula for valuable comments about a draft of 
this letter and also T. Yumibayashi for informing him of the web site \cite{Fields Medal Symposium} 
of Fields Medal Symposium.



\begin{thebibliography}{99}
\bibitem{BPST} A.~A.~Belavin, A.~M.~Polyakov, A.~S.~Schwarz, Yu.~S.~Tyupkin, Phys. Lett. B59 (1975) 85.  
\bibitem{BogoPS} E.~B.~Bogomolnyi, Sov.~J.~Nucl.~Phys.~24 (1976) 449; 
M.~K.~Prasad, C.~M.~Sommerfield, Phys.~Rev.~Lett.~35 (1975) 760.
\bibitem{AHDM} M.~F.~Atiyah, N.~J.~Hitchin, V.~G.~Drinfeld, Yu.~I.~Manin, 
Phys. Lett. A65 (1978) 185.  
\bibitem{Nahm80} W.~Nahm, Phys. Lett. B90 (1980) 413. 
\bibitem{Nahm82} W.~Nahm, The construction of all self-dual multimonopoles by 
the ADHM method, Monopoles in Quantum Field Theory, in: N.~S.~Craigie, P.~Goddard, W.~Nahm (Eds.), 
Proc. Monopole Meeting, Trieste Italy, December 1982, World Scientific.
\bibitem{Hit86} N.~J.~Hitchin, The Self-duality equations on a Riemann surface. Proc. London Math. Soc. 55 (1987) 59.
\bibitem{Lohe} M. A.~Lohe, Phys. Lett. B70 (1977) 325.
\bibitem{CorriGodd} E.~Corrigan, P.~Goddard, Ann.~Phys.~154 (1984) 253.
\bibitem{Fields Medal Symposium} FIELDS MEDAL SYMPOSIUM, http://www.fields.utoronto.ca/programs/\\
scientific/fieldsmedalsym/12-13/program.html, (2012). 
\bibitem{COFN} E.~Corrigan, D.~I.~Olive, D.~B.~Fairlie, J.~Nuyts, Nucl.~Phys.~B106 (1976) 475.
\bibitem{tH} G.~'t\,Hooft, Nucl.~Phys.~B79 (1974) 276.
\bibitem{Polyakov} A.~M.~Polyakov, JETP Lett.~20 (1974) 194.
\bibitem{MantonSut} N.~Manton, P.~Sutcliffe, Topological Solitons, Cambridge Monographs on Mathematical Physics, Cambridge, 2004.
\bibitem{CherkisKap} S.~Cherkis, Kapustin, Commun. Math. Phys. 218 (2001) 333.
\bibitem{NO} H.~B.~Nielsen, P.~Olesen, Nucl.~Phys.~B61 (1973) 45.
\bibitem{Kamata1978} M.~Kamata, Progr. Theor. Phys.~59 (1978) 1346.
\bibitem{Dirac monopole} P.~A.~M.~Dirac, Phys. Rev.~74 (1948) 817.
\bibitem{AtiyahWard} M.~F.~Atiyah, R.~S.~Ward, Commun. Math. Phys.~55 (1977) 117.
\end{thebibliography}
\end{document}